\def\td{{\rm d}}

\documentclass[twocolumn,preprintnumbers,amsmath,amssymb,aps,floatfix,eps,superscriptaddress]{revtex4}

\usepackage{graphics}	% Include figure files
\usepackage{bm}		% bold math
\usepackage{alltt}	% verbatim-like typewriter environment

\begin{document}

\title{Direct conversion of rheological compliance measurements into storage and loss moduli}

\author{R. M. L. Evans}\email{mike.evans@physics.org}
\affiliation{School of Physics and Astronomy, University of Leeds, LS2 9JT, U.K.}

\author{Manlio Tassieri}\email{m.tassieri@elec.gla.ac.uk}
\affiliation{Department of Electronics and Electrical Engineering, University of Glasgow, G12 8LT, U.K.}

\author{Dietmar Auhl}
\affiliation{School of Physics and Astronomy, University of Leeds, LS2 9JT, U.K.}

\author{Thomas A. Waigh}
\affiliation{School of Physics and Astronomy, University of Manchester, Oxford  
Rd, M13 9PL, UK.}

\date{12 December, 2008}

\begin{abstract}
We remove the need for Laplace/inverse-Laplace transformations of experimental data, by presenting a direct and straightforward mathematical procedure for obtaining frequency-dependent storage and loss moduli ($G'(\omega)$ and $G''(\omega)$ respectively), from time-dependent experimental measurements. The procedure is applicable to ordinary rheological creep (stress-step) measurements, as well as all microrheological techniques, whether they access a Brownian mean-square displacement, or a forced compliance. Data can be substituted directly into our simple formula, thus eliminating traditional fitting and smoothing procedures that disguise relevant experimental noise.
\end{abstract}


\maketitle

The procedure that has become established \cite{Bird87}, for obtaining frequency-dependent dynamic moduli from non-oscillatory rheometry, is to fit the experimental data to a particular model (often the generalized Maxwell model is used), and subsequently to calculate the resulting complex viscoelastic modulus, $G^*(\omega)\equiv G'(\omega) + i G''(\omega)$,
for that parametrization of the model. That procedure can be somewhat restrictive, as it may force the user to approximate their data into the prescribed form, or to use a very large number of fitting parameters. It also artificially hides experimental noise, making the uncertainties in the final results difficult to quantify. Equivalently, one can find an approximate Laplace transform of the time-dependent data \cite{Mason97}, then derive the Laplace transform of the stress relaxation modulus, and subsequently transform from a Laplace to a Fourier description (either numerically or, for certain functional forms, analytically). In either case, the procedure limits the user's freedom in the types of formulae that can easily be fitted and manipulated, and it can be somewhat laborious. Furthermore, in cases where  the experimental data are imperfectly fitted by the pre-conceived functions, or other approximations are introduced (such as an approximate Laplace transform \cite{Mason97} or its inverse \cite{contin}), the accuracy of the derived moduli becomes vague. Here, we show that a more direct, straightforward and accurate treatment of rheometric data is possible, and derive a formula for $G^*(\omega)$ in terms of the experimental data points themselves.

Let us begin by summarising why the conversion of time-dependent rheometry data into storage and loss moduli is traditionally such a complicated process. (For more discussion and approximate solutions, see \cite{Schwarzl70,Tschoegl89}.)
In principle, a simple relationship exists between the dynamic moduli, $G'(\omega)$ and $G''(\omega)$, and the experimentally accessible time-dependent compliance, $J(t)$. In a shear-creep (or stress-step) experiment, 
\begin{equation}
\label{defineJ}
  J(t)=\gamma(t)/\sigma_0
\end{equation}
is the ratio of the time-dependent shear strain $\gamma(t)$ to the magnitude $\sigma_0$ of the constant stress that is switched on at time $t=0$. Alternatively, in a passive microrheology experiment, the function $J(t)$ is proportional to the mean square displacement of a probe particle \cite{Mason97,Xu} executing Brownian motion in the viscoelastic fluid (averaged over many trajectories). Since the relaxation modulus is related to the compliance by a convolution \cite{Ferry80},
\begin{equation}
\label{convolution}
	\int_0^\tau G(t)\,J(\tau-t)\,\td t = \tau,
\end{equation}
it is, in principle, a simple matter to extract the modulus by deconvolving Eq.(\ref{convolution}), using an integral transform, such as the Fourier transform. So the Complex viscoelastic modulus $G^*(\omega)$ (which is the Fourier transform of the time derivative of $G(t)$), is a simple function of the Fourier-transformed compliance:
\begin{equation}
\label{G*}
	G^*(\omega) = \frac{1}{i\omega \widehat{J}(\omega)}
\end{equation}
where $i\omega\widehat{J}\equiv J^*$ is sometimes called the dynamic compliance \cite{Ferry80}.
The problem that arises is due to the fact that the Fourier transform (denoted \mbox{${\cal F}[\ldots]\!(\omega)$ )} of the compliance,
\begin{equation}
\label{divergent}
	\widehat{J}(\omega) \equiv {\cal F}[J]\!(\omega)
	\equiv \int_{-\infty}^\infty J(t)\, e^{-i\omega t}\,\td t,
\end{equation}
is not a convergent integral, since $J(t)$ grows with increasing time.
(Even for a solid, where $J(t)$ tends to a finite constant at long times, the integral in Eq.~(\ref{divergent}) remains undefined.) This has led investigators to resort instead to a Laplace transform, $J_L(s) \equiv \int_0^\infty J(t)\, e^{-s t}\,\td t$, since its integral is convergent. Deconvolution of Eq.~(\ref{convolution}) thus yields $G_L(s)$, the Laplace tranform of $G(t)$. Only if the resulting function is expressed as a simple formula, then the desired complex modulus can finally be obtained by analytic continuation, $G^*(\omega) = \left. s\,G_L(s) \right|_{s=i\omega}$.

In this Letter, we discuss how to bypass the above foray into Laplace space, by working directly with the Fourier transform of the compliance. Although its integral representation in Eq.~(\ref{divergent}) is not convergent, the quantity $\widehat{J}(\omega)$ is nonetheless well defined, as is apparent in Eq.~(\ref{G*}), since $G^*(\omega)$ exists for all real finite $\omega$. It is well known how to find the Fourier transform of an unbounded function such as $J(t)$. We shall nevertheless introduce the method in pedagogical detail, in order to clarify its applicability to any causality-respecting compliance function, including that dictated by the raw experimental data. This will allow us to find a simple formula for the viscoelastic moduli that are implied by the data. 

\begin{figure}
	\centering
		\resizebox{86mm}{!}{\includegraphics{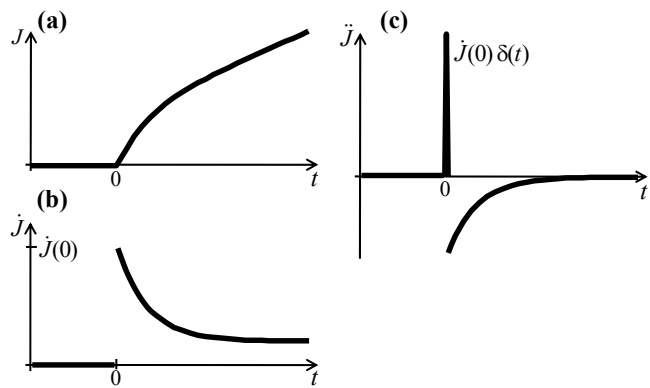}}
	\caption{\label{JFig}(a) Sketch of a typical time-dependent compliance, $J(t)$, which must vanish for negative $t$ due to causality. (b) Its first derivative, $\dot{J}(t)$. (c) The second derivative, $\ddot{J}(t)$.}
\end{figure}

First we note that causality requires
\begin{equation}
\label{causality}
	J(t)=0 \qquad \mbox{for}\qquad t<0
\end{equation}
as there can be no response before the stress step is applied. So $J(t)$ is a function resembling the sketch in Fig.~\ref{JFig}a, and $\widehat{J}(\omega)$, required in Eq.~(\ref{G*}), is its Fourier transform. In the long-time limit, the response of a fluid (viscoelastic or otherwise) to an imposed step stress is to undergo shear at a constant rate. So that the compliance $J(t)$, which is proportional to the total strain, asymptotes to a straight line (see Fig.~\ref{JFig}a) with a gradient equal to the reciprocal of the static viscosity. So $\ddot{J}(t)$, the second derivative of $J(t)$, is a function that vanishes at large $t$,
and {\em its} Fourier transform therefore converges. We can reconstruct the former Fourier transform from the latter, since they are simply related thus:
\begin{equation}
\label{related}
	\widehat{J}(\omega) = \frac{-1}{\omega^2} \: {\cal F}[\ddot{J}\,]\!(\omega).
\end{equation}
It is not immediately obvious that a function can be reconstructed from its second derivative, in this way, as there is a danger of losing information about the absolute offset and slope of the original function $J(t)$. However, we can retain that information since we have, as a reference, the known part of the function given in Eq.~(\ref{causality}). To do so, we must perform the double differentiation of $J(t)$ over the {\em whole} of the function's domain, including negative and zero values of $t$, as shown schematically in Fig.~\ref{JFig}b and c. Notice that $\dot{J}(t)$ has a discontinuity at $t=0$, of size $\dot{J}(0)$, the initial gradient (at time $t=0^+$) of the compliance. Differentiating the discontinuous function $\dot{J}(t)$ yields the second derivative, $\ddot{J}(t)$ which vanishes at negative times, has a Dirac delta function of strength $\dot{J}(0)$ located at $t=0$, and is finite for positive $t$ (see Fig.~\ref{JFig}c). Substituting that form into Eq.~(\ref{related}) yields
\begin{equation}
\label{intermediate}
	\widehat{J}(\omega) = \frac{-\dot{J}(0)}{\omega^2} \;
	- \; \frac{1}{\omega^2} \int_{0^+}^\infty  e^{-i\omega t}\, \ddot{J}(t)\, \td t,	
\end{equation}
where the integration is over positive values of $t$ only, for which the function $\ddot{J}(t)$ is non-singular. That integration can be performed by standard Fourier transform methods if we define a simpler function $J_2(t)$ that is the second derivative of $J(t)$ excluding the delta function,
\begin{equation}
	J_2(t)	\equiv	\left\{
	\begin{array}{ll}
		\ddot{J}(t)	&	\qquad \mbox{for} \qquad t>0,	\\
			&	\\
		0	&	\qquad \mbox{otherwise.}
	\end{array}
	\right.
\end{equation}
In terms of $J_2(t)$, the integral in Eq.~(\ref{intermediate}) is simply 
${\cal F}[J_2]\!(\omega)$, so that Eq.~(\ref{G*}) yields a simple formula for the storage and loss moduli,
$G'(\omega)+iG''(\omega)=i\omega/\{\raisebox{0pt}[10pt]{$\dot{J}$}(0) 
+{\cal F}[J_2]\!(\omega)\}$.

Let us generalize to include compliance functions that are discontinuous at $t=0$, since a finite discontinuity $J(0)\equiv\lim_{t\to0^+}J(t)\neq0$ is often observed, reflecting the fact that the data-acquisition rate cannot access the regime of a material's response preceeding a small-$t$ plateau in $J$. Such a compliance function has a delta-function contribution to its {\em first} derivative $\raisebox{0pt}[10pt]{$\dot{J}$}(t)$, of strength $J(0)$. Hence, defining $J_1(t)$ to exclude that delta function, so that $\dot{J}(t)=J_1(t)+J(0)\,\delta(t)$, we have
$\widehat{J}(\omega) = \{J(0)+{\cal F}[J_1](\omega)\}/i\omega$,
ultimately yielding
\begin{equation}
\label{result2}
	G'(\omega) + i G''(\omega) = \frac{i\omega}
	{i\omega J(0) + \raisebox{0pt}[10pt]{$\dot{J}$}(0) + {\cal F}[J_2](\omega)}.
\end{equation}

\begin{figure}
	\centering
		\resizebox{40mm}{!}{\includegraphics{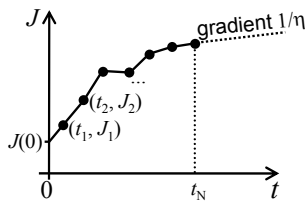}}
	\caption{\label{piecewise} An example of data points $(t_k,J_k)$ from experimental measurements of the compliance as a function of time. The data are interpolated by a piecewise-linear function, and also linearly extrapolated to infinity by a final line segment of gradient $\eta^{-1}$, which vanishes in the case of elastic solids for which the present treatment is equally valid.}
\end{figure}

Since Eq.~(\ref{result2}) holds for any compliance function, we now apply it directly to the experimental data, by defining the piecewise linear function $J(t)$ that interpolates between data points, depicted in Fig.~\ref{piecewise}. It is obvious that the resulting angularity and non-monotonicity of the function are unphysical consequences of the experimental noise, but attempting to remove them by using a smooth fitting function would constitute doctoring the data. Instead, we take the data at face value by using this function that passes through every data point. Since evaluation of the Fourier transform requires $J(t)$ to be defined for all positive $t$, while experimental data are finite, our piecewise linear function must include an extrapolation to $t=\infty$, thus introducing one extra parameter, the steady-state viscosity $\eta$ (see Fig.~\ref{piecewise}). This is not a peculiarity of the present method; {\em any} data-analysis for converting $J(t)$ to $G^*(\omega)$ requires such extrapolation, though some methods obscure it in arcane algorithms. The second derivative of our experimental function $J(t)$ is a series of delta functions, so its Fourier transform is trivial to evaluate. The strengths of the delta functions are equal to the discontinuities in gradient of $J(t)$, and Eq.~(\ref{result2}) becomes straighforwardly expressed in terms of the experimental data points $(t_i,J_i)$, which need not be equally spaced:
\begin{eqnarray}
\label{result3}
  \frac{i\omega}{G^*(\omega)} &=&  i\omega J(0)
  + \left( 1-e^{-i\omega t_1}\right) \frac{\left(J_1-J(0)\right)}{t_1}
  + \frac{e^{-i\omega t_N}}{\eta}	  \nonumber	\\
  &+& \sum_{k=2}^N \left( \frac{J_k-J_{k-1}}{t_k-t_{k-1}} \right)
  \left( e^{-i\omega t_{k-1}}-e^{-i\omega t_k} \right)  
\end{eqnarray}
Note that Eq.~(\ref{result3}) requires the convention $t_1>0$ 
so that, if the compliance is non-zero at $t=0$, it enters the formula in the value of $J(0)$ only.

Equation~(\ref{result3}) makes a direct link between the experiment and the resulting graphs of $G'$ and $G''$, thus removing any subjective judgement from the results, and allowing genuine experimental noise and uncertainties to appear on those graphs for critical evaluation. The formula also has the advantage of being very easy to apply. For instance, the code required to evaluate it using Mathematica$^\circledR$ can be written in just three lines, as follows\footnote{Similar implementations using Maple$^\circledR$ and MatLab$^\circledR$ can be found at www.pcf.leeds.ac.uk/research/highlight/view/4}.
\begin{alltt}
\{t,J\}=Transpose[Import["filename.txt","Table"]];

G[\(\omega\)_, J0_, \(\eta\)_, \{t_, J_\}] := I \(\omega\)/(I \(\omega\) J0 + 
  (1-Exp[-I \(\omega\) t[[1]]])(J[[1]]-J0)/t[[1]] + 
  Exp[-I \(\omega\) t[[Length[t]]] ] / \(\eta\) + 
  Sum[ (Exp[-I \(\omega\) t[[k-1]]]-Exp[-I \(\omega\) t[[k]]])
  (J[[k]]-J[[k-1]])/(t[[k]]-t[[k-1]]), 
  \{k,2,Length[t]\} ] );

LogLogPlot[\{Re[G[\(\omega\),0.0000023,1145300,\{t,J\}]], 
Im[G[\(\omega\),0.0000023,1145300,\{t,J\}]]\},\{\(\omega\),0.001,1000\}]                  
\end{alltt}
(Some installations of Mathematica$^\circledR$ first require a library to be loaded, using 
\texttt{<< Graphics`Graphics`}.) The first line imports the experimental data (a list of pairs of numbers) from a file (here named \texttt{filename.txt}). The second line defines the function in Eq.~(\ref{result3}), and the third displays the resulting real and imaginary parts of the complex modulus, in this case using the parameter values $J(0)=2.3\times10^{-6}$, $\eta=1.1453\times10^6$ that characterise our data.

\begin{figure}
	\centering
		\resizebox{85mm}{!}{\includegraphics{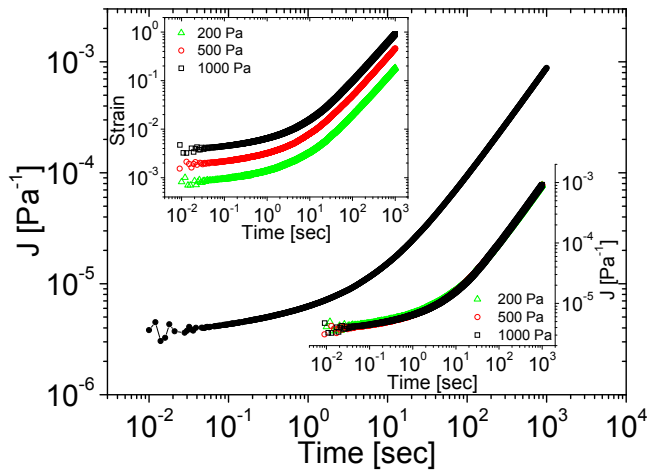}}
	\caption{\label{compliance} Creep compliance $J$ \textit{vs}.~time $t$, for PI ($M_{w} = 150000$, $M_w/M_n=1.03$) measured at $T=0^{o}C$. {\bf Top left inset:} strain \textit{vs}.~time for three different applied stresses of magnitude $200$~Pa (green triangles), $500$~Pa (red circles) and $1000$~Pa (black squares). {\bf Bottom right inset:} compliance \textit{vs}.~time, for the same data sets. The data-collapse testifies to the linearity of the creep measurements.}
\end{figure}

We have used Eq.~(\ref{result3}) to obtain the frequency-dependent storage and loss moduli of a near-monodisperse polyisoprene melt with a weight-average molar mass $M_w$ of 152 kg/mol and polydispersity $M_w/M_n$ of 1.03 (where $M_n$ is number-average molar mass). The time-dependent creep compliance $J(t)$ was determined in a stress-step measurement performed at 0 Celsius using a commercial AntonPaar MCR-501 rheometer (cone diameter 25mm, cone angle 1 degree).
The insets of Fig.~\ref{compliance} show the measured strain curves (top left inset), and thus the compliance curves (bottom right inset), at three different shear stresses of magnitude 200~Pa, 500~Pa and 1000~Pa. The good superposing of the three compliance curves indicates that the applied stresses were small enough to access the fluid's linear regime (where the ratio on the RHS of Eq.~(\ref{defineJ}) is independent of $\sigma_0$), but large enough to provide a satisfactory signal-to-noise ratio. 

\begin{figure}
	\centering
		\resizebox{70mm}{!}{\includegraphics{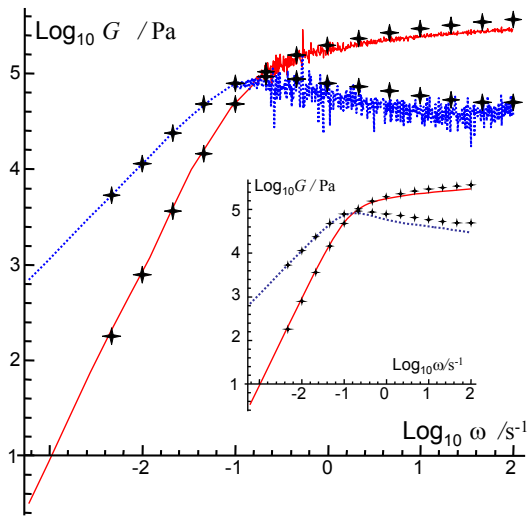}}
	\caption{\label{ModulusFig} Storage (red solid line) and loss (blue dotted line) moduli, respectively the real and imaginary parts of $G^*$ found by substituting compliance data (Fig.~\protect\ref{compliance}) into Eq.~(\protect\ref{result3}). Crosses are data from oscillatory rheometry of the same fluid, shown for comparison. {\bf Inset:} To demonstrate the efficacy of Eq.~(\protect\ref{result3}) for smooth functions also, the compliance data (Fig.~\protect\ref{compliance}) were fitted to the function $J(t)=(t/\eta)+a+b \tanh\left(c+d \ln t+e \ln^2 t+f \ln^3 t\right)$, the form of which is motivated only by fit quality. The seven parameter values were $\eta=1.1453\times10^6$Pa, $a=4.7\times10^{-6}$Pa$^{-1}$, $b=2.4\times10^{-6}$Pa$^{-1}$, $c=0.288082$, $d=0.247501$, $e=0.0174205$, $f=0.000685812$, yielding $J(0)=2.3\times10^{-6}$Pa$^{-1}$. The result of substituting 200 sampled points from this function, at equal logarithmic-time intervals, into Eq.~(\protect\ref{result3}) are plotted in the inset, with the oscillatory data again reproduced for comparison.}
\end{figure}

The data points in Fig.~\ref{compliance} (main graph) are the average of three compliance measurements. Scatter due to experimental noise is apparent, particularly at small $t$, but these raw data were substituted directly into Eq.~(\ref{result3}) without smoothing or fitting. The resulting functions $G'(\omega)$ and $G''(\omega)$ are plotted in Fig.~\ref{ModulusFig} (red and blue curves respectively). Notice that the curves are not smooth. This demonstrates a virtue of our straightforward deconvolution method (Eq.~(\ref{result3})) over established methods: that it preserves the experimental noise. The noise visible in the curves is a true reflection of the experimental uncertainties.

For comparison, Fig.~\ref{ModulusFig} also shows data from oscillatory measurements on the same fluid. The agreement is good across five orders of magnitude in modulus and in frequency. Not only are features such as the characteristic relaxation times (e.g.~where the moduli cross) accurately obtained, but absolute values of the moduli are consistent, up to random errors in experimental reproducibility, thus demonstrating that the same linear rheological information can be extracted from the stress-step experiment as from a large number of oscillatory experiments.

The moduli from Eq.~(\ref{result3}) have been plotted in Fig.~\ref{ModulusFig} for angular frequencies $\omega$ in the domain $\omega_{\rm min}<\omega<\omega_{\rm max}$. Outside this frequency window, the moduli given by Eq.~(\ref{result3}) are dominated by artifacts. The {\em lowest} accessible frequency, $\omega_{\rm min}\approx t_N^{-1}$, is determined by the experiment's duration $t_N$. Obtaining a single data point by oscillatory rheometry at that same frequency $\omega_{\rm min}$ would require several complete oscillations, thus taking an order of magnitude longer than the creep measurement of the entire dynamic spectrum.
The {\em highest} accessible frequency, $\omega_{\rm max}\approx t_1^{-1}$, is set by the early-time resolution of the stress-step experiment, where the first reliable data are obtained at time $t_1$. At high frequencies, however, oscillatory measurements are relatively quick to perform and can yield better precision than can be obtained from the short-time creep response of the rheometer (depending on details of the instrument's design). Hence, in practice, using a combination of creep and oscillatory measurements may be the best strategy to determine a fluid's entire dynamic spectrum.

Finally, we note that Eq.~(\ref{result3}) has another use besides substitution of raw data. Even for analytical functions $J(t)$ (such as might be used in theoretical work, or to approximate noisy data), standard numerical algorithms can fail to evaluate the Fourier transform required in Eq.~(\ref{result2}), if the small-$t$ behaviour of $J$ is non-trivial. Equation~(\ref{result3}) turns out to be a reliable method for numerically evaluating the required function, with accuracy greatly superior to simple quadrature algorithms such as trapesium rule. To demonstrate this, a smooth function (see caption to Fig.~\ref{ModulusFig}) was fitted to the compliance data in Fig.~\ref{compliance}, and was then sampled at logarithmically-uniform intervals in $t$, for substitution into Eq.~(\ref{result3}). This yielded the smooth curves in the inset to Fig.~\ref{ModulusFig}. We found convergence on the exact result for 200 sample points, and that the above Mathematica$^\circledR$ code executed in a matter of seconds, whereas the built-in \texttt{FourierTransform} algorithm stalled when applied to evaluate Eq.~(\ref{result2}).

In summary, dynamic moduli can be straightforwardly obtained by substitution of compliance data into Eq.~(\ref{result3}), which is equally valid for viscoelastic fluids or solids. The equation is quick to evaluate, removes the need for approximate fitting or obscure black-box algorithms, and correctly preserves the experimental noise that is so crucial to good scientific methodology.

{\small ACKNOWLEDGMENTS: We are grateful to Alexei Likhtmann for informative discussions. RMLE is funded by the Royal Society. MT, DA and TAW were funded by EPSRC.}

\vfill

\bibliography{rheo}

\end{document}